# Irradiation-Induced Spin Bath Evolution and As-Grown Hydrogen Defects in CVD Diamond Revealed by NV-Based DEER Spectroscopy

*Olga Rubinas,[1,2,*] Jeroen Prooth,[1,2] Michael Petrov,[1] Remy Vandebosch,[1] Emilie Bourgeois,[1,2] David Chvátil,[3] Milos Nesladek[1,2]*

[1]*Institute for Material Research (IMO), Hasselt University, Wetenschapspark 1, Diepenbeek, Belgium*
[2]*IMOMEC, IMEC, Kapeldreef 75, Heverlee, Belgium*
[3]*Department of Accelerators Nuclear Physics Institute of the CAS, Rez Husinec 25068, Czech Republic*

Corresponding author*: milos.nesladek@uhasselt.be

**Abstract: The aim of this paper is to provide the reader with review and current state of the art of fabrication high T2 coherence diamond, optimised by the use of double electron-electron resonance (DEER) spectroscopy. Using DEER we study formation, transformation, and annealing paramagnetic defects in as grown CVD diamond and after post-processing. Electron irradiation leads to the formation of an additional S=1/2 resonance in the DEER spectrum, which we consider to be a composite X ensemble. By tracking the concentrations of X and P1 during annealing from 650 to 1200 °C, we find that the X ensemble initially consists of a mixture of $V^-$ spins and interstitial spins, which disappear at about 650 °C. Vacancies migrate during annealing, forming clusters that persist to ~1000 °C and disappear upon annealing at 1200 °C, contributing to the X ensemble signal. We have also developed a model of the influence of the mixed spin bath on the coherence of NV centers, which includes independent couplings with P1 centers, $V^-$, divacancies, and interstitials. Detailed DEER studies allowed us to reveal and resolve a weak signal from two additional S=1/2 species associated with hydrogen: $NVH^-$ and consistent with substitutional hydrogen defect, which overlap the vacancy spectral line. Taken together, these results show that the NV-DEER method is a powerful tool for investigating paramagnetic defects in diamond with high precision and nanoscale resolution, essential for material optimisation. The achieved high T2 coherence time is consistent with the spin bath model, and the crystals reach the quality required for advanced quantum sensing applications.`**

## I. INTRODUCTION

Nitrogen-Vacancy (NV) centers in diamond have long been studied and employed as sensors for a wide range of applications, including precision magnetometers[1,2] and gyroscopes [3,4], as well as in nanoscale nuclear magnetic resonance (NMR) spectroscopy[5], biological and temperature sensing[6]. The most common route to fabricate such sensors is chemical vapor deposition (CVD) growth[7] with controlled nitrogen incorporation, followed by post-processing, i.e., high energy irradiation and annealing[8] to increase the conversion of nitrogen into $NV^-$ centers. Electron[9], neutron[10], or proton[11] irradiation creates vacancies in the crystal lattice, which migrate during annealing to form $NV^-$ centers. However, particle irradiation damages the lattice and degrades spin properties; for example, one study reported a broadening of the NV-center spin-resonance linewidth with increasing electron-irradiation dose during post-processing[12]. Although this workflow is well studied, sensor development typically focuses on the properties of the diamond after annealing [13]. By contrast, the characteristics of the material and the NV ensemble immediately after irradiation have received much less attention. However, the defects present in diamond after the growth can be also considered as precursor for the final material quality. To address this issue we focus on all of the NV fabrication stages, especially on the intermediate stage, examining the changes in diamond defects immediately after irradiation and their systematic transformation during annealing. Rather than optimizing a specific irradiation or annealing protocol, the goal of this work is to identify the defect species that constitute the spin bath at each stage of NV fabrication. By resolving the microscopic composition and spin dynamics of the defect environment, our study establishes a framework for future optimization of CVD diamond growth and post-processing strategies based on the actual spin-bath landscape revealed here, extending beyond the conventional consideration of substitutional nitrogen alone[14]. Because spin coherence and as a result resonance linewidths are determined by interactions with this surrounding defect bath, understanding the spin environment is necessary for improving the performance of NV ensembles. Finally, we achieve $T_2$ times that reach the spin bath limit, and lead to material that is ready to use in quantum sensing applications.

Early photoluminescence (PL) studies demonstrated the creation of optically active centers such as GR1 (neutral vacancy, $V^0$), ND1 (negatively charged vacancy, $V^-$)[15,16], and interstitial-related defects, including 3H[17], during the irradiation process. These works established and identified their optical transitions and annealing behavior. Some of these defects have electron spins that affect via the spin bath the spin properties of $NV^-$ centers and limit their coherence and subsequently sensing applications[18]. Early electron paramagnetic resonance (EPR) studies identified $V^-$ (ND1) as an $S = 3/2$[19] center and reported additional paramagnetic centers such as nickel (HPHT diamond)-[20], phosphorus-[21], and nitrogen-related clusters[22].

The conventional method for studying paramagnetic centers in solids is electron paramagnetic resonance (EPR)[23]. Owing to its limited spin sensitivity, conventional X-band CW EPR spectrometers[24] typically requires on the order of $10^{12}$-$10^{13}$ spins, i.e., tens of ppb to a few tenths of a ppm when averaged over a millimeter-scale diamond sample. In addition, classical EPR provides no intrinsic spatial resolution, as the signal is integrated over the entire sample volume. In contrast, double electron-electron resonance (DEER) spectroscopy based on the coherence of NV centers[25] combines high spin sensitivity[26] with spatial selectivity enabled by confocal optical addressing. The probed volume is defined by the confocal excitation volume of the NV centers (typically on the order of a few cubic micrometers), allowing local measurements of spin concentrations and enabling the investigation of individual layers of multilayer CVD diamond samples. Importantly, NV-based DEER enables direct quantification of paramagnetic spin densities without requiring calibration to an external reference standard. Single-NV implementations of DEER have further demonstrated the capability to detect individual proximal paramagnetic spins via dipolar coupling[27]. Previously, DEER in diamond has been used to quantify substitutional nitrogen (P1 centers)[28] and to detect the $NVH^-$ defect[29]. Recently, DEER has also emerged as a powerful tool for characterizing some of paramagnetic defects in diamond before and after NV centers' formation during annealing process[30].

In this work, we employ DEER spectroscopy[31,32] as the main tool to characterise the defects across the three stages of defect formation: before irradiation; after irradiation, but before annealing; and after annealing. The main aim of the paper is to establish an experimentally determined and DEER-monitored sequence of paramagnetic defect transformations during CVD diamond post-processing. Specifically, our work tracks how CVD growth conditions and subsequent electron irradiation introduce a vacancy/interstitial-related paramagnetic ensemble, how this ensemble further evolves upon annealing, and how these changes correlate with the NV-center coherence, with specific emphasis on the formation of newly monitored defects. In particular, we focus on the negatively charged vacancy $V^-$ (ND1) and report the first DEER detection of resonance consistent with substitutional hydrogen in diamond[33]. A Hamiltonian-based model is employed to interpret the DEER spectra and to account for contributions from vacancies, $NVH^-$, and other paramagnetic species.

We studied a set of CVD laboratory-grown diamond samples, using MW-PECVD, with various nitrogen concentrations. Further on we fabricated staggered diamond layer and structures, fabricated from one deposition run that permit to study on the same samples the mechanism of defect and NV generation without being influenced by the condition and reproducibility of the growth apparatus such as vacuum conditions before the growth, leak rates etc. To this end we prepared following sampled: S1 (1–4 ppm, four nitrogen-doped layers staggered in a vertical structure, separated by nitrogen-free buffers), S2 (~10 ppm, single nitrogen-doped layer), S3 (25–40 ppm, four nitrogen-doped layers, staggered in a vertical structure, separated by buffers). These samples deferred for example by the growth temperature that is essential for defining the growth mechanism and N uptake and NV/N formation during the CVD diamond growth. DEER spectroscopy was applied to these samples to detect paramagnetic defects before and after electron irradiation. We also used a commercial CVD monolayer sample S4 (~1 ppm initial nitrogen) as a reference sample. For the reference sample S4, we tracked the complete sequence from electron irradiation through stepwise annealing between 650 °C and 1200 °C, monitoring the formation, transformation, and annihilation of paramagnetic centers at each stage. Additional irradiation-induced defects not associated with nitrogen were detected, revealing their influence on $NV^-$ center coherence. We compared the $T_2$ coherence time at all measurement stages and assessed the contribution of various paramagnetic defects formed by post-processing to the NV centers' coherence. We studied not only the defects induced by post-processing but also H-related defects which are formed during the CVD growth.

Thus, NV-DEER enables quantitative analysis of paramagnetic defects in CVD diamonds at the nanoscale, including less noticeable species such as substitutional hydrogen. This method allowed us to track how defects arise and transform during irradiation and annealing, as well as how they affect the spin properties of NV centers.

## II. RESULTS AND DISCUSSION

### A. CVD growth and post-processing

Three types of CVD diamond layers were grown for this study on commercially available type-Ib HPHT substrates cut from a single Sumitomo crystal and polished by Almax Easylab (samples S1, S2, and S3). Two of the samples (S1 and S3) were designed as multilayers composed of alternating nitrogen-free (buffer) and nitrogen-doped layers (L1–L4) (Figure 1a), while S2 contained only one buffer and one nitrogen-doped layer. The multilayer architecture was chosen to enable controlled variation of the growth temperature within a single diamond sample, allowing systematic investigation of its effect on nitrogen incorporation and NV formation while minimizing sample-to-sample variability. In particular, if the study is conducted in several runs, the reproducibility of conditions is not guaranteed. For example, the history of the reactor walls strongly influences the growth process, and similarly, the position of the sample relative to the plasma also affects the crystal growth and settings for temperature measurements. Therefore, multilayer growth offers significant advantages when comparing the spin properties of the samples.

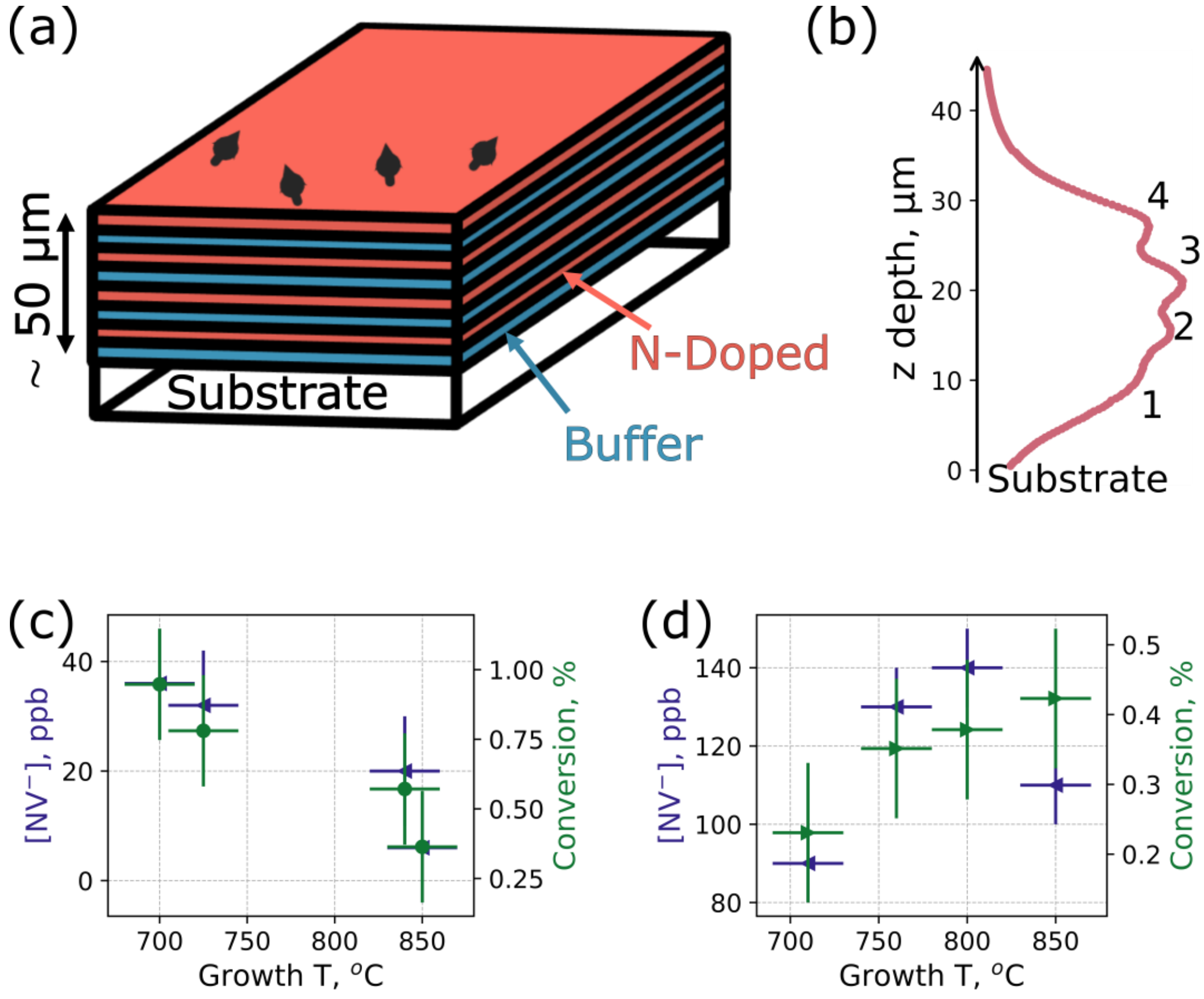

**Figure 1**. Growth design and nitrogen incorporation trends in CVD diamond samples. (a) Schematic of the multilayer diamond structure used for samples S1 and S3, composed of alternating nitrogen-free (buffer) and nitrogen-doped layers (L1–L4) grown on type-Ib HPHT

substrates. (b) Confocal z-scan photoluminescence of a representative multilayer sample showing clear optical contrast between nitrogen-free and nitrogen-doped layers, confirming successful layer stacking. (c) Growth temperature dependence of NV incorporation and $NV/N_S$ conversion efficiency for the low-nitrogen sample S1. (d) Growth temperature dependence for the high-nitrogen sample S3, demonstrating the opposite trend.

Before deposition, all substrates were cleaned in a standard hot $H_2SO_4$ : $KNO_3$ acid mixture for several hours and ultrasonicated twice in ultrapure water to remove graphitic and amorphous carbon residue from the diamond surface. Each sample was then loaded into the CVD reactor and subjected to a 20-min oxygen plasma etch at 100 Torr in a 2% $O_2$: $H_2$ mixture to prepare the surface for growth. Oxygen plasma treatment was used to remove the surface polishing-induced damage, etch the non-diamond carbon impurities as well as the diamond material close to dislocations teaching the surface. We optimized this technique previously allowing reaching the near atomically flat surfaces[34].

The diamond buffer layers were then deposited at 140 Torr using a 3% $CH_4$: $H_2$ gas mixture with the substrate temperature maintained at 880°C. The nitrogen-doped layers were grown under the conditions summarized in Table I.

**Table I**. Growth conditions of samples.

| | $[N_2]$, sccm | $[H_2]$, sccm | $[CH_4]$, sccm | T, °C |
|---|---|---|---|---|
| S1(L1) | 0.072 | 480 | 20 | 700 |
| S1(L2) | 0.072 | 480 | 20 | 720 |
| S1(L3) | 0.072 | 480 | 20 | 840 |
| S1(L4) | 0.072 | 480 | 20 | 850 |
| S2 | 0.012 | 99 | 1 | 800 |
| S3(L1) | 0.072 | 97 | 3 | 710 |
| S3(L2) | 0.072 | 97 | 3 | 760 |
| S3(L3) | 0.072 | 97 | 3 | 800 |
| S3(L4) | 0.072 | 97 | 3 | 850 |

Growth parameters were selected to study how temperature and nitrogen concentration affect substitutional-nitrogen ($N_S$) and NV center incorporation and their spin properties. The relative nitrogen fraction in the gas phase compared to $CH_4$ was approximately 3600 ppm for S1, 12 000 ppm for S2, and 24 000 ppm for S3. The total thickness of each structure was kept below 50 μm (Figure 1b) to ensure optimal microwave coupling during spin-resonance. Microwave accessibility across all layers is confirmed by nearly identical π-pulse durations (within ≈10%) at fixed microwave power.

All nitrogen-containing layers exhibited ensembles of $NV^-$ centers with concentrations between 10–50 ppb, determined via the photon count-rate saturation method[35]. After synthesis, all samples underwent initial characterization by DEER spectroscopy to assess the influence of gas composition and growth temperature on the P1 center concentration and $NV^-$ conversion efficiency.

Interestingly, S1 and S3, grown at different N concentration ranges, exhibited opposing trends in NV incorporation and $NV/N_S$ conversion ratio as a function of growth temperature (Figure 1c,d). In S1, the NV incorporation rate decreased with increasing temperature, consistent with previous reports[36]. In contrast, S3, grown under much higher nitrogen concentration, displayed an enhanced conversion efficiency, suggesting a different temperature dependence at high nitrogen partial pressure than it was not reported previously. Changes in growth kinetics with nitrogen addition were observed in earlier studies[37], indicating that both temperature and nitrogen fraction strongly influence nitrogen incorporation mechanisms. Following the initial measurements, all samples were electron-irradiated at an energy of 7 MeV with a dose of $10^{18} e/cm^2$, at a temperature below 40°C. Post-irradiation characterization was performed using the same set of optical and spin-resonance techniques to evaluate changes in defect structure and electronic properties.

A commercially available monolayer diamond type IIa CVD diamond from Evolve Diamond (S4) served as the reference sample in this work. S4 contained 1 ppm P1 centers and ≈ 10 ppb $NV^-$ centers, and no other detectable impurities, as verified by Photoluminescence (PL) and DEER spectra. The reference sample S4 was used as received and did not undergo any irradiation or annealing treatment prior to this study. All samples were tested before and after irradiation, which showed no significant changes in NV content. Here, "NV content" refers to the optically detected $NV^-$ signal under our readout conditions (continuous optical excitation), which may differ from the dark charge-state equilibrium after irradiation. To study thermal evolution of the NV centers conversion and other properties (Section B, C), the reference diamond S4 was subsequently annealed stepwise at 650 °C, 850 °C, 1000 °C, and 1200 °C for 2h at each temperature in a home-built high-vacuum chamber. $NV^-$ centers concentration was tested on each step (Table II) and revealed significant enhancement at 1200 °C. Such behavior observed at higher annealing temperatures (≥1000 °C) is consistent with vacancy release from vacancy clusters and stabilization of the $NV^-$ charge state, as reported in recent high-temperature annealing studies[38–41].

**Table II**. $NV^-$ centers concentration of the S4 sample vs annealing temperature.

| Annealing T | 0 °C | 650 °C | 800 °C | 1000 °C | 1200 °C |
|---|---|---|---|---|---|
| $[NV^-]$, ppb | 11±5 | 30±5 | 38±10 | 50±20 | 200±20 |

## B. Irradiation-related defects in DEER and PL spectroscopy

### *1. Photoluminescence spectroscopy*

To identify optical centers and monitor defect evolution, PL spectroscopy was used first alongside DEER measurements. The as-grown CVD samples exhibited only the characteristic $NV^-$ emission with the zero-phonon line (ZPL) at 637 nm and $NV^0$ emission with the ZPL at 575 nm[42] (Figure 2a, b). No other visible defect-related luminescence was detected within the 550–900 nm range.
After electron irradiation, an additional ZPL at 741 nm appeared (Figure 2b) in the low-nitrogen samples (S1, S2, and the reference S4). This emission corresponds to the GR1 center (the neutral vacancy $V^0$) in diamond[43]. The appearance of GR1 confirms the successful creation of vacancies and provides a convenient optical marker for irradiation-induced lattice damage. Its presence in samples with low P1 concentration (1-10 ppm) indicates a shortage of available electron donors, preventing full conversion of $V^0$ into $V^-$. This does not imply the absence of negatively charged vacancies, but rather the coexistence of $V^0$ and $V^-$ charge states. In contrast, the high-nitrogen sample S3 ([P1] ≈30 ppm) showed no GR1 emission (Figure 2a), consistent with efficient vacancy charging due to donor availability.

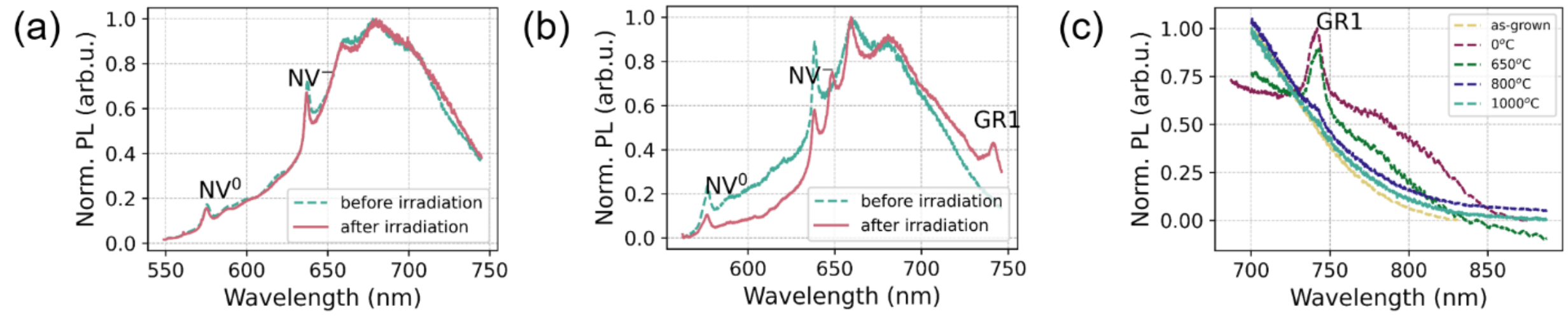

**Figure 2.** Photoluminescence spectra of CVD diamond samples before and after irradiation and during annealing. (a) The high-nitrogen sample S3(L4) ([P1] ≈ 30 ppm) was recorded before and after electron irradiation. No GR1 signal appears. (b) PL spectra of the medium-nitrogen sample S2 ([P1] ≈ 10 ppm) before and after irradiation, showing the emergence of the GR1 center's ZPL at 741 nm, characteristic of neutral vacancies ($V^0$). (c) Evolution of the PL spectrum of the low-nitrogen reference sample S4 ([P1] ≈ 1 ppm) through successive stages: as-grown, after irradiation, and following annealing at 650–1000 °C.

Upon thermal annealing, the GR1 intensity gradually decreased and disappeared after treatment at 1000 °C (Figure 2c), consistent with the migration and aggregation of vacancies into divacancy complexes[44].

## *2. DEER methodology*

DEER is a pulsed magnetic-resonance technique that quantitatively probes dipolar coupling between two classes of paramagnetic spins, denoted here as *A* (probe spins) and *B* (pumped spins). In DEER, the *A* spins generate a spin-echo, while a resonant π-pulse applied to the *B* spins perturbs their magnetic dipole field, producing a modulation of the echo amplitude. This modulation encodes the dipolar interaction strength, spatial distribution of *B* spins, and thus their concentration. For two electron spins separated by a vector r, the secular dipolar interaction is[23]:

$$\hat{H}_{dip} = \frac{\mu_0}{4\pi}\frac{\mu_B^2 g_A g_B}{r^3}(1 - 3\cos^2\theta)\hat{S}_Z^A\hat{S}_Z^B, \tag{1}$$

with θ is the angle relative to the external field. When a π-pulse flips a *B* spin, the local dipolar field at an *A* spin changes sign, shifting the accumulated phase during the spin-echo evolution. For an *A*-spin echo sequence of length 2τ, the accumulated dipolar phase is

$$\delta\varphi = \frac{\mu_0}{4\pi}\frac{\mu_B^2 g_A g_B}{\hbar}(1 - 3\cos^2\theta)\frac{2T}{r^3}\sigma_B, \tag{2}$$

where T is the DEER pump-pulse delay, and $\sigma_B = \pm 1/2$ is the *B*-spin projection[32]. Averaging over a three-dimensional spin bath[45] with density n yields an exponential DEER decay. Following previous work, the normalized DEER intensity is

$$I_{DEER}(T) = \exp\left[-\frac{2\pi\mu_0\mu_B^2 g_A g_B}{9\sqrt{3}\hbar} nT \langle \sin^2\frac{\theta}{2}\rangle_L\right], \quad (3)$$

where the factor $\langle \sin^2\frac{\theta}{2}\rangle_L$ accounts for the finite excitation bandwidth via convolution with the spin-resonance line shape L(f). Equation (3) provides a direct quantitative link between the experimentally observed DEER contrast and the B-spin concentration.

The probability that a *B* spin with detuning Δ is flipped by a pump pulse of duration $t_B$ is given by the Rabi formula:

$$P_R(\Delta, t_B) = \frac{\Omega^2}{\Omega^2+\Delta^2}\sin^2\left[\frac{1}{2}\sqrt{\Omega^2+\Delta^2}t_B\right], \quad (4)$$

where Ω is the resonant Rabi frequency. The effective excitation of the entire *B*-spin ensemble is the convolution:

$$P_B(f_B, t_B) = \int_{-\infty}^{+\infty} L(\xi)\, P_R(f_B - \xi, t_B)d\xi. \quad (5)$$

Substituting Equation (5) into the exponent of Equation (3) yields the general DEER formula including spectral broadening:

$$I_{DEER}(f_B, t_B, T) = \exp\left[-\frac{2\pi\mu_0\mu_B^2 g_A g_B}{9\sqrt{3}\hbar} nTP_B(f_B, t_B)\right]. \quad (6)$$

This expression makes explicit that the DEER response is governed by the spectral tuning of the pump-pulse frequency, the duration of the inversion pulse, and its temporal placement within the echo sequence, as well as by the density of the target spins that contribute to the dipolar field.

In the context of the theoretical framework above, the nitrogen–vacancy (NV) center in diamond provides a practical realization of the *A* spin. Its electronic spin can be optically initialized and read out, enabling precise detection of the dipolar-field changes induced by selectively driving the surrounding paramagnetic impurities, which act as the *B* spins. When these neighboring spins (most commonly P1 centers or other defect-related species) are inverted by a resonant microwave pulse, the resulting change in their dipolar field leads to a measurable modulation of the NV spin-echo amplitude. Interpreting this modulation through the DEER formalism allows one to extract the spectral characteristics and local concentration of the *B*-spin ensemble, thereby revealing the nanoscale magnetic environment that determines NV coherence in diamond. One limitation of the DEER measurements is that only B-spins with sufficiently long $T_2$ to be coherently driven by the RF pulses contribute to the signal; very short-lived spins may therefore be underrepresented.

### *3. DEER spectroscopy in diamond*

Building on this methodology, we applied DEER spectroscopy to directly detect and quantify paramagnetic defects created during sample processing. The pulse sequence was based on a Hahn-echo protocol (Figure 3a) with an additional π-pulse applied at the resonance frequency of the target spin ensemble. This pulse flips the spins of the target species, modifying the local magnetic field experienced by the $NV^-$ probe and producing a measurable drop in its echo amplitude. This method allows investigating samples with high spatial resolution[46] which is necessary for multilayer samples in our case.

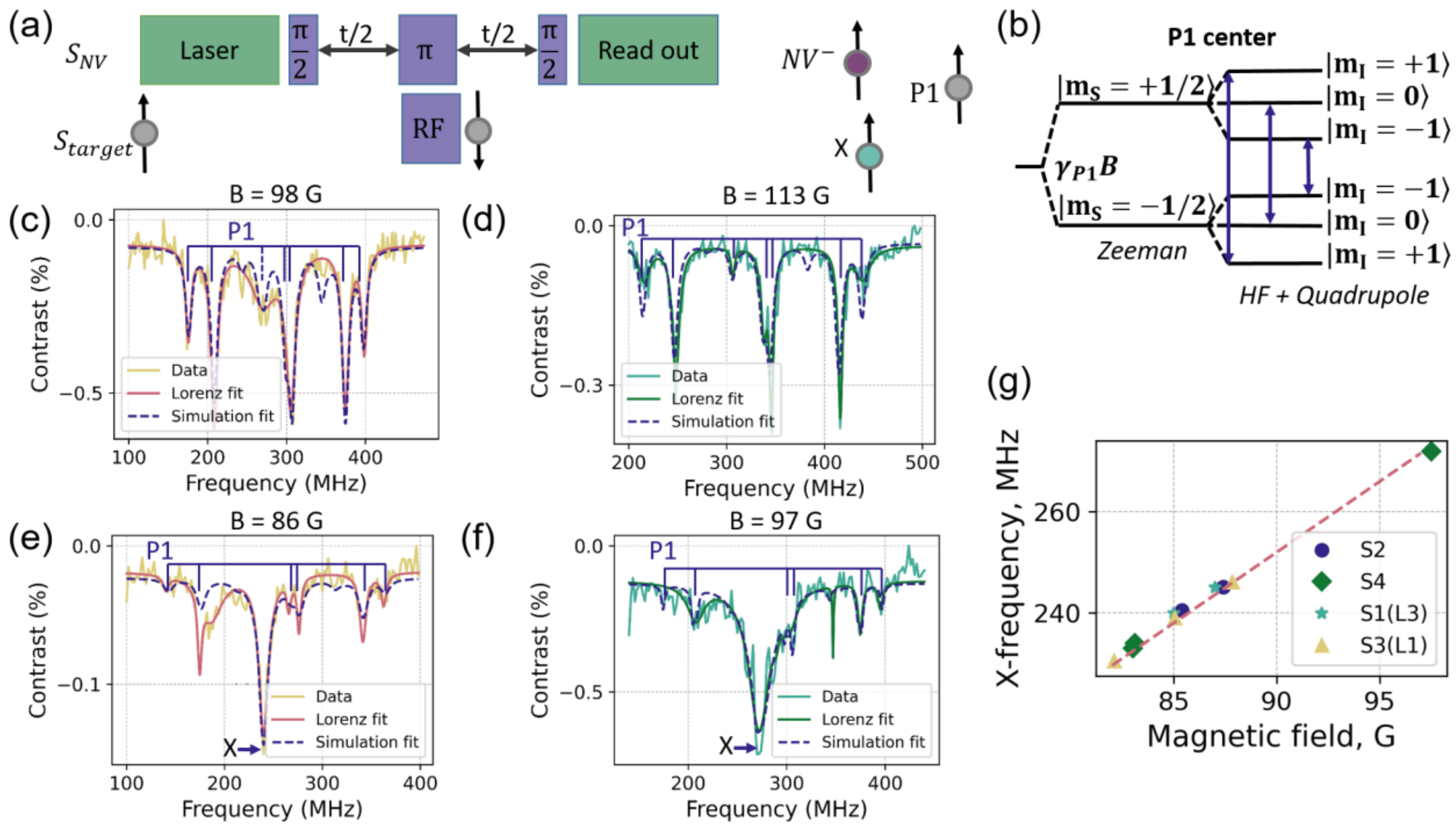

**Figure 3.** DEER spectroscopy of CVD diamond samples before and after irradiation. (a) Pulse sequence of the double electron–electron resonance (DEER). (b) Energy-level diagram of the P1 center ($S = \frac{1}{2}, I = 1$) showing the three allowed hyperfine transitions. (c) DEER spectrum of the laboratory-grown sample S1(L1) ([P1] = 3.8 ppm) before irradiation, exhibiting only the P1 transitions, allowed as well as forbidden. (d) DEER spectrum of the reference diamond S4 ([P1] = 1 ppm) before irradiation, exhibiting only the P1 transitions, allowed as well as forbidden. (e) DEER spectrum of S1(L1) after electron irradiation, revealing the new resonance X. An additional intermittent feature below ~185 MHz is observed in Figure 3e. As it does not correspond to any known defect transitions and is not reproducible across measurements, it is attributed to an experimental artifact and excluded from the quantitative analysis. (f) DEER spectrum of S4 after irradiation, likewise showing a dominant X-line and suppressed P1 intensity. (g) Position of the X-center resonance as a function of magnetic field. The linear dependence (yellow dashed line) with slope 2.8 MHz/G confirms the $S = ½$ nature of the X ensemble.

Before irradiation, all samples exhibited the characteristic P1 center transitions (Figure 3b) arising from substitutional nitrogen ($N_S^0, S = \frac{1}{2}, I = 1$)[47]. These transitions were modeled using the established spin Hamiltonian for the P1 defect:

$$H_{P1}/h = \frac{g_e \mu_B}{h} \vec{B} g \vec{S} + \frac{\mu_N}{h} \vec{B} \vec{I} + \vec{S} A \vec{I} + \vec{I} Q \vec{I}, \quad (7)$$

where A and Q are isotopically symmetrical matrices with $A_\parallel = 114$ MHz, $A_\perp = 81$ MHz, $Q_\perp = -3.97$ MHz. Because of symmetry the quadrupole interaction is simplified as $Q_\perp {S_z}^2$.

Simulated spectra using these parameters (blue dashed lines in Figure 3c,d) accurately reproduce the experimentally observed triplet pattern corresponding to the $m_I = +1, 0, -1$ transitions of P1 centers. One of the important issues in this model is the external magnetic field ($\vec{B}$). P1 centers have $C_{3v}$ symmetry because of the Jahn-Teller effect[48] and are distributed along all four diamond axes equally as $NV^-$-centers. In this case, we need to calculate energy levels for all four projections of the magnetic field. To make this easier, we always aligned the B-field along one of the diamond axes and have only two projections due to symmetry. That is why the spectrum of the P1 center contains 6 resonances (three for each projection). Magnetic field alignment along a single ⟨111⟩ axis was verified for each DEER measurement using NV ODMR (see Supplementary Materials).

To calculate transition probabilities between spin states, we included the microwave (MW) interaction in the system's Hamiltonian and applied the rotating wave approximation (RWA)[49]. Specifically, the total Hamiltonian was constructed as the sum of the ground-state spin Hamiltonian $H_{gs}$ (which is $H_{P1}$ here and $H_X$, $H_{V^-}$, $H_{NVH^-}$ or $H_H$ further in the text) and the MW interaction term $H_{MW}$. The general form of the total Hamiltonian is:

$$H = H_{gs} + H_{MW} = H_{gs} - \gamma_e \vec{S}\vec{b}_{MW} \cos 2\pi f_{MW} t, \tag{8}$$

where $\vec{b}_{MW}$ and $f_{MW}$ are the amplitude and frequency of the oscillating microwave field, respectively. To evaluate transition probabilities, the total Hamiltonian was transformed into the eigenbasis of $H_{gs}$ using a unitary transformation U. This yielded the transformed operators:

$$\begin{gathered} H^0_{gs} = U^\dagger H_{gs} U, \\ H^0_{MW} = U^\dagger H_{MW} U, \\ H^0 = H^0_{gs} + H^0_{MW} \end{gathered} \tag{9}$$

The resulting operator $H^0$ describes transitions between the system's eigenstates. The RWA was then applied to the matrix elements $h_{ij}$ of $H^0$, allowing us to ignore rapidly oscillating terms and keep only those near resonance. Transitions were considered allowed when the matrix element $|h_{ij}| > 10^{-1}$. For these transitions, the coupling strength was estimated from the Rabi frequency as: $F_{ij} \approx |h_{ij}|/2\hbar \approx \Omega_R/2$. In our model, we also accounted for the angles of the MW and external field as fit parameters.

In addition to the expected P1 resonances, samples (S1–S3 in all layers) showed a broad, weak feature near 275 MHz at a field of ≈98 G Figure 3c). It could be caused by the forbidden transition of P1 center as it correlates with the Hamiltonian-based fit model (Figure 3c). However, it might also be due to hydrogen related defects as grown in our diamonds as they have transitions in the same frequency ranges (see Supplementary Materials).

The reference sample S4 revealed only P1 center's peaks in DEER spectrum with no extra features. This observation does not imply the absence of hydrogen in the sample, but rather indicates that hydrogen-related paramagnetic centers are below the detection limit of our DEER measurements or exist in magnetically inactive configurations. Therefore, all resonances observed in this sample before irradiation can be fully described by the P1 Hamiltonian plus a minor contribution from a forbidden transition at 304 MHz, that was covered by model-fit, with no evidence of vacancy-type centers at this stage (Figure 3d).

After electron irradiation, a new strong resonance appeared near the center of the DEER spectra in all samples (Figure 3e,f). The corresponding frequency shift with magnetic field follows the electron gyromagnetic ratio $\gamma = 2.8\ \mathrm{MHz/G}$, confirming an $\mathrm{S} = ½$ center (Figure 3g). The peak is therefore assigned to defects that are related to irradiation with g ~ 2 that we denote as X-centers, rather than to a single species. We discuss the X-defects later in the text. The spectra can be quantitatively described by a combined spin-Hamiltonian model including both P1 and X transitions:

$$\mathrm{H_{P1+X}}/\mathrm{h} = \mathrm{H_{P1}}/\mathrm{h} + \frac{g_e \mu_B}{h} \vec{B} g \vec{S}, \qquad (10)$$

where we assumed $\mathrm{S} = 1/2$ and $\mathrm{g} = 2.0027$. A g-factor was taken for the negatively charged vacancy $\mathrm{V^-}$ in previous works[19]. From this work, it is known that the actual electron spin of the $\mathrm{V^-}$ defect is $\mathrm{S} = 3/2$. However, as noted there, its EPR response is effectively indistinguishable from that of a spin-1/2 system. Because the defect has $\mathrm{T_d}$ symmetry, the zero-field splitting vanishes, and all allowed transitions collapse into a single line that behaves as an effective $-1/2 \leftrightarrow 1/2$ transition. For this reason, in the present analysis, we treat $\mathrm{V^-}$ as an effective spin-1/2 center. Simulated spectra using these parameters reproduce the full experimental line shape: the P1 triplet transitions, plus the new central X-line. This confirms that the observed spectra arise from the coexistence of P1 centers and an additional paramagnetic ensemble with a nearly free-electron g-factor.

Our study shows that attributing X resonance solely to $\mathrm{V^-}$ is incompatible with the full temperature-dependent behavior under annealing, discussed below. To examine this further, we analyzed the defect evolution in the reference sample S4 where no H-related defect was present. In the as-irradiated state, the expected donor balance [$\mathrm{P1_{before}} = \mathrm{P1_{after}} + \mathrm{X}$] is not satisfied: although S4 initially contained 1 ppm of P1 centers, the DEER measurement yielded 3.7 ppm of X centers after irradiation. Even under the hypothetical assumption that X corresponds exclusively to $\mathrm{V^-}$, applying the required factor of 1/2 to convert the effective spin-$1/2$ DEER concentration to the actual density of an $\mathrm{S} = 3/2$ vacancy defect: $[\mathrm{V^-}]_{\mathrm{true}} = \frac{1}{2}[\mathrm{X}]_{\mathrm{DEER}}$ (see Supplementary Material for details of the averaging correction) still yields ≈1.8 ppm of $\mathrm{V^-}$, far exceeding the available donor population. Thus, the X resonance cannot originate from $\mathrm{V^-}$ alone.

Additional evidence arises from the annealing behavior. Upon heating to 650 °C, the X concentration drops to approximately 1 ppm. This temperature range is in the range or above the well-established annealing of interstitial-related defects[50]. While isolated interstitial defects in diamond[51] are generally known to exhibit $\mathrm{S} = 1$ (with high symmetry) and no established $\mathrm{S} = 1/2$ interstitial centers have been reported, the dramatically decreasing of the X resonance at the characteristic interstitial-annealing temperature[52] introduces an apparent inconsistency: X displays a clear $\mathrm{S} = 1/2$ signature, yet its thermal behavior tracks that of interstitial-related defects. This mismatch indicates that X cannot correspond to a simple interstitial. Instead, the data are more consistent with X being a composite or complex defect that includes an interstitial component or is structurally related to an interstitial configuration which spin concentration is altered by annealing at the mentioned concentration, probably forming more complex structures around vacancies. This point is not extensively studied in the literature. Taken together, the imbalance in paramagnetic centers accounting, the necessary vacancy-spin correction, and the annealing behavior all indicate that X is not a pure vacancy defect but instead a composite signal arising from a mixture of $V^-$ and interstitial-related contributions.

Photoluminescence data support this interpretation: at 650 °C, the GR1 ($V^0$) line remains strong, implying that most vacancies were not annealed out. At 800 °C, the DEER-derived X concentration remains roughly constant (~1 ppm), but the GR1 emission nearly disappears, consistent with the formation of divacancies $V_2^-$[53] (or other V-related complexes [43]) that contribute to the same $S = 1/2$ resonance as a part of $S = 3/2$ spectrum[44]. At 1000°C, the concentration of X-defect started to decrease to 0.7 ppm. Finally, after 1200 °C, the X-signal crucially reduced to the P1's contrast level (Figure 4a), but was still present in the spectra. Its concentration was estimated as 0.1 ppm, matching the known temperature of vacancy clusters $V_n^-$ (>1200 °C). These correlations demonstrate that the post-irradiation DEER resonance is a superposition of several vacancy- and interstitial-related centers evolving sequentially with thermal treatment:

$$0\ °C: (V^- + I) \rightarrow 650\ °C: (V^-) \rightarrow 800\ °C: (V^- + V_2^-) \rightarrow 1000\ °C: (V_2^- + \text{low } V_n^-) \rightarrow 1200\ °C: (\text{low } V_n^-).$$

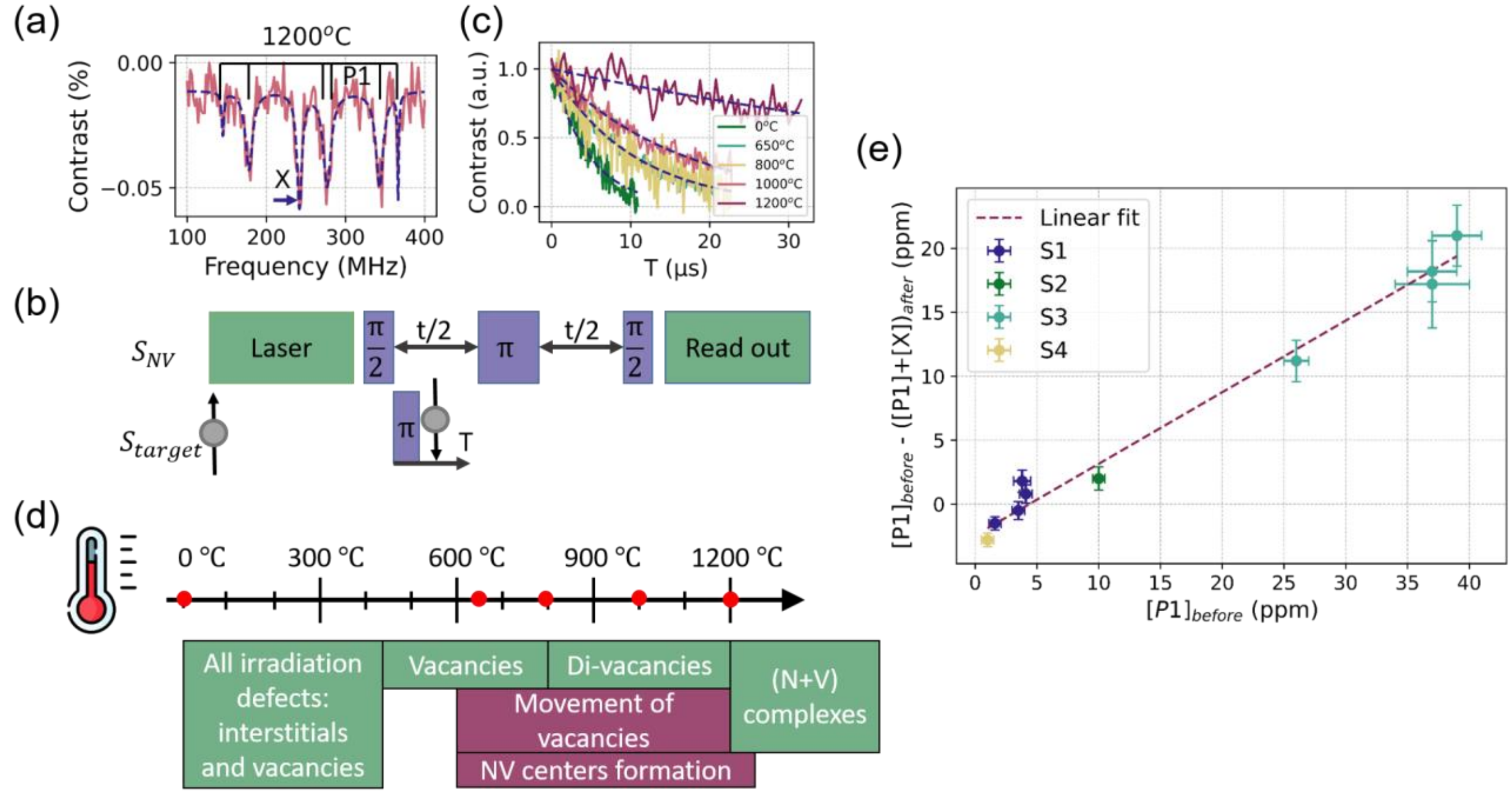

**Figure 4.** Evolution of X-center and donor concentrations revealed by DEER spectroscopy. (a) DEER spectrum of sample S4 after 1200 °C annealing, showing that the X-center resonance remains of comparable amplitude to the P1 center transitions, indicating the persistence of a small residual paramagnetic population. (b) DEER-decay pulse sequence used to quantify spin-bath concentrations and dipolar coupling strengths between NV centers and surrounding paramagnetic defects. (c) DEER decay curves on X resonance of sample S4 recorded after successive annealing steps from as irradiated to 1200 °C, demonstrating the progressive reduction of the X-center signal. (d) Schematic qualitative illustration of the dominant defect populations and processes inferred

from the DEER and PL measurements at different annealing stages. Green boxes demonstrate defects the diamond contains; Violet boxes shows processes during annealing; Red points represent temperatures in our study. (e) The loss of the total number of paramagnetic defects (detected in DEER) shows a nearly linear scale with the nitrogen level (P1 centers) before irradiation.

In all these concentration measurements, a modified DEER pulse scheme was used. By varying the delay of the π-pulse (Figure 4b), we obtained DEER-decay curves from which the average dipolar coupling (Equation (6)) and hence the spin concentration were extracted. We measured X center's content evolution by this method (Figure 4c). To aid the discussion, we include a schematic summary of the dominant defect species and processes inferred from our DEER and PL data across the annealing sequence (Figure 4d). This schematic reflects the experimentally observed trends and annealing thresholds, rather than a detailed microscopic defect census.

Another observation is that in all samples, the P1 concentration dropped after irradiation, even in high-nitrogen material, indicating partial transfer of an electron to another defect (for example all in X resonance) and transition to an interstitial state[54]. Moreover, in the intermediate-nitrogen sample (S2) and the high-nitrogen sample (S3 – with all layers), the sum of the measured P1 and X concentrations remained below the initial P1 level (Table III), suggesting additional "P1 losses". That is opposite to samples with low nitrogen content (S1(L4), S4), where the paramagnetic center grew after irradiation. We considered this decrease in P1 centers and revealed that the dependence of the "P1 losses" = difference between initial P1 content and the sum of P1 centers after irradiation and X concentrations. Interestingly, the "P1 losses" depend on the initial P1's content across samples, showing an approximately linear correlation (Figure 4e), i.e. higher the initial P1 concentration is, the spin loses are higher after post-processing by annealing, supporting a nitrogen-driven redistribution mechanism. We excluded isolated nitrogen interstitials that were reported in previous work[55] as they have S = ½, are visible in EPR spectra with hyperfine splitting, and must be distinguished in our DEER spectra. From all these facts, we can conclude the formation of previously theoretically predicted diamagnetic nitrogen-interstitial pairs[56,57] or non-activated previously experimentally observed nitrogen-vacancy[58] pairs, non-visible under our DEER conditions.

**Table III.** Concentration before and after irradiation of spin defect in our samples.

| | [P1] before irradiation, ppm | [P1] after irradiation, ppm | [X] after irradiation, ppm | $[NV^-]$ before irradiation, ppb | $[NV^-]$ after irradiation, ppb |
|---|---|---|---|---|---|
| S1(L1) | 3.8±0.7 | <<1 | 1.1±0.1 | 39±10 | 56±20 |
| S1(L2) | 4.1±0.5 | <<1 | 2.3±0.1 | 35±10 | 54±20 |
| S1(L3) | 3.5±0.5 | <<1 | 3.0±0.1 | 22±10 | 17±10 |
| S1(L4) | 1.6±0.5 | <<1 | 3.0±0.1 | 7±3 | 10±3 |
| S2 | 10±0.5 | 6.0±0.7 | 2.0±0.3 | 1±0.5 | 9±3 |
| S3(L1) | 39.0±2.0 | 14±0.8 | 4.0±1.0 | 90±20 | 100±20 |
| S3(L2) | 37.0±3.0 | 12±1.3 | 7.8±1.0 | 130±20 | 130±20 |
| S3(L3) | 37.0±2.0 | 12±1.3 | 6.5±1.0 | 140±20 | 150±20 |
| S3(L4) | 26.0±1.0 | 11±1.3 | 4.0±1.0 | 110±20 | 120±20 |
| S4 (ref) | 1.0±0.5 | <<1 | 3.7±0.5 | 11±5 | 13±5 |

For comparison, MeV electron irradiation of diamond is known to produce vacancy concentrations on the order of 0.1–5 ppm, as reported in optical absorption studies[59]. The X-center concentrations observed here are consistent with this range, while reflecting the presence of multiple irradiation-induced defect species.

### C. Coherence time analysis

The spin coherence time $T_2$ of the NV ensemble was investigated in all samples (S1 – S4) in both cases: just after growth and after electron irradiation. After irradiation, we studied coherence dynamics followed by stepwise annealing of the commercial reference crystal from Evolve diamonds (S4). All samples had a natural $^{13}$C abundance 1.1%. $NV^-$, $NV^0$, $NVH^-$ concentration was in the range of << 1ppm, so we won't consider them in coherence properties analysis.

Before irradiation, all samples were tested for $T_2$ by Hahn-echo[60] decay of $NV^-$ centers (Figure 5a,b), and demonstrated $T_2$ scaling inversely with the P1 concentration (Figure 5c). The dependence is well described by a dipolar spin-bath model in which decoherence arises from magnetic noise generated by the P1 electron spins[18]:

$$\frac{1}{T_2} = a[P1] + \frac{1}{T_{2,(13C)}}. \quad (11)$$

Here a is the dipolar coupling constant and $T_2$, $^{13}C \approx 700$ μs which represents the upper limit imposed by the $^{13}$C nuclear-spin bath[61]. Fitting yields $a = (5.5 \pm 0.3) \times 10^{-3} (\text{ppm} \cdot \mu\text{s})^{-1}$, in good agreement with previous reports for high-purity CVD diamond[18]. This confirms that decoherence in untreated samples is determined exclusively by the P1 spin bath.

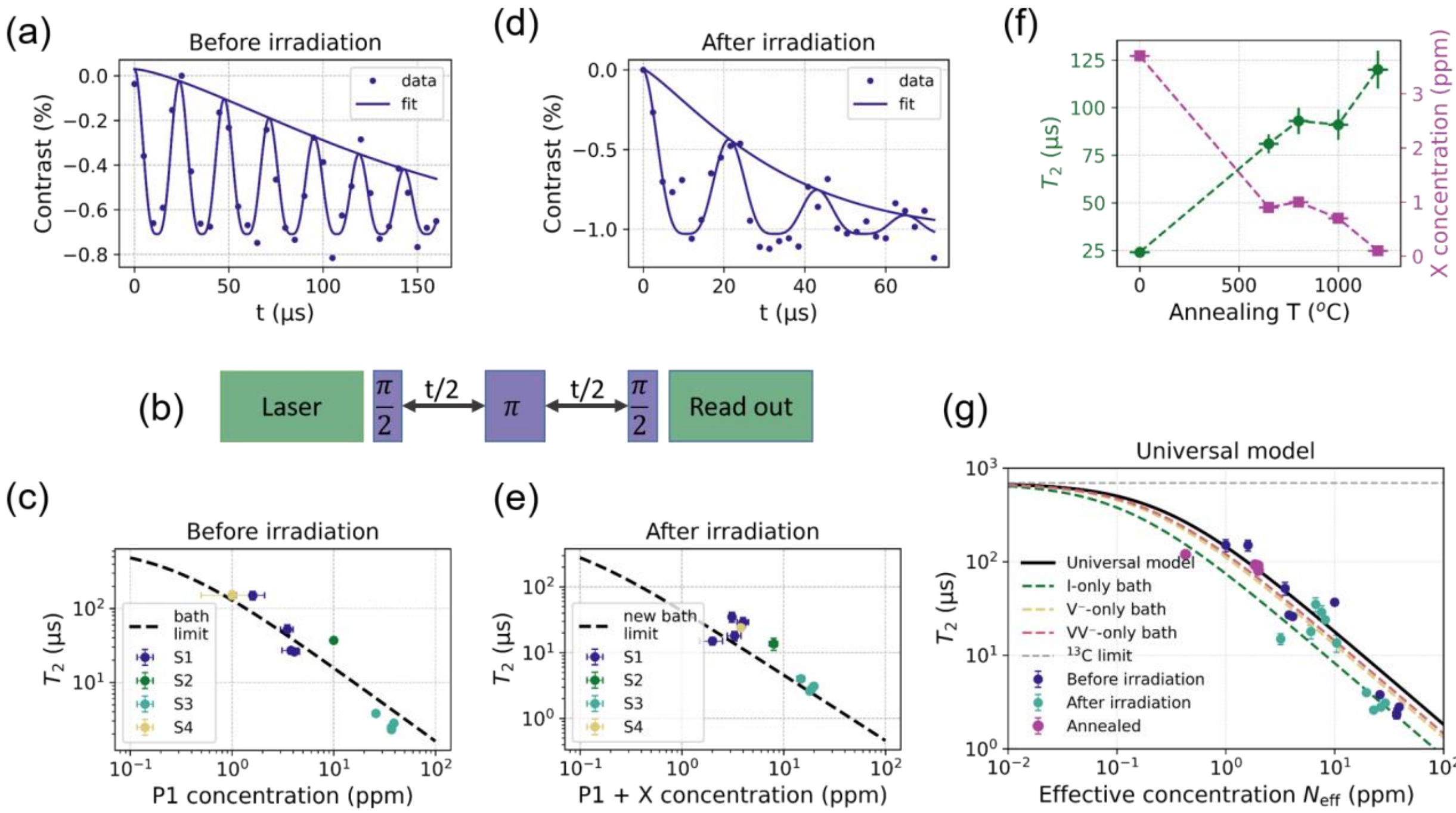

**Figure 5.** Spin coherence properties of $NV^-$ ensembles in CVD diamond before and after irradiation. (a) Hahn-echo decay measured on the reference sample S4 (before irradiation). (b) Pulse sequence of the Hahn-echo and DEER protocols used for $T_2$ characterization. (c) Correlation of the ensemble $T_2$ with P1 concentration in as-grown samples, together with the classical dipolar-bath model. (d) Hahn-echo decay of the sample S4 after irradiation. (e) Dependence of $T_2$ on the combined defect concentration $[P1] + [X]$ after irradiation, fitted with the corrected spin-bath model including the composite X-ensemble ($V^-$ + interstitial-related defects). (f) Variation of $T_2$ (green symbols) and X-center concentration (pink symbols) as a function of annealing temperature for sample S4. (g) Universal model describing all $T_2$ results from as-grown, irradiated, and annealed samples using the combined bath expression. S1 S2, S3, S4 – names of samples, used in this paper (see Section A).

Following electron irradiation, the P1 concentration decreased markedly in all samples, accompanied by the appearance of new paramagnetic centers (Section B). Simultaneously, the NV-ensemble coherence time collapsed (Figure 5d), confirming that the irradiation-induced centers act as an additional magnetic-noise bath. The dependence of $T_2$ on total defect density $[P1] + [X]$ is well described by an extended model (Figure 5e):

$$\frac{1}{T_2} = a[P1] + e([X]) + \frac{1}{T_{2,(13C)}}, \tag{12}$$

where X denotes the ensemble of irradiation-induced defects. A linear fit yields $e = (2.1 \pm 0.5) \times 10^{-2} (\mathrm{ppm} \cdot \mu\mathrm{s})^{-1}$. As discussed in Section B, the X-ensemble includes both negatively charged vacancies $V^-$ and all interstitial-related defects "I", so that $[X] = [V^-] + [I]$. The coefficient e, therefore, represents the effective decoherence strength of this mixed defect population. Note that in the dipole model, we do not include the concentration of hydrogen related defects, discussed in Section D, as their contribution to the decoherence will be minimal compared to other spin defects, in particular X defects.

To resolve the individual contributions, we used the annealing experiment on sample S4, where both $T_2$, and the concentrations of all paramagnetic species were measured at each step. Immediately after irradiation, $T_2$ shortened from 150 μs to ~25 μs. After annealing at 650 °C, $T_2$ the increases to ~80 μs, while the X-center intensity decreases from 3.7 ppm to ~1 ppm, consistent with the removal of interstitial defects at this temperature[50]. As we discussed in Section B, the real concentration of $V^-$ is two times lower than the DEER-detected one. Using the corrected concentration of $V^-$ at 650 °C, $T_2$ at that step, and a basic equation analogous to (12), we can extract the decoherence coefficient for $V^-$: $b = \frac{1/T_2 - 1/T_{2,(13C)} - a[P1]}{[V^-]} = (1.5 \pm 0.4) \times 10^{-2} (\mathrm{ppm} \cdot \mu s)^{-1}$. This value is significantly larger than that for P1 centers, which may be attributed to the higher spin multiplicity of $V^-$ defects, S = 3/2, resulting in stronger dipolar coupling to NV centers.

To quantify the contribution of interstitial-related defects to decoherence, we first estimate the relative fractions of $V^-$ and interstitials in the as-irradiated S4 sample. DEER decay measurements yield a total X-center concentration of 3.7 ppm in the just-irradiated state. After annealing at 650 °C, the X signal is reduced to ~1.0 ppm and is dominated by $V^-$ centers, consistent with the removal of interstitial-related defects at this temperature. Assuming that the number of $V^-$ centers does not change significantly between irradiation and the first anneal, as $V^0$ centers in ZPL spectra changed for 11% only (see Section B), we take $[V^-]_{irr} = 1.1$ ppm) and therefore assign the remaining $[I]_{irr} = 2.6$ ppm to interstitial-related spins. Using the extended decoherence model (12) and the previously determined $V^-$ coefficient b, we separate the effective post-irradiation slope e into vacancy and interstitial contributions and estimate interstitial-related defects decoherence: $d = \frac{1/T_2 - 1/T_{2,(13C)} - a[P1] - b[V^-]}{[I]} \approx (1.2 \pm 0.3) \times 10^{-2} (\mathrm{ppm} \cdot \mu s)^{-1}$. The comparatively large value of d may be partly overestimated, since a fraction of the interstitial-related defects are DEER-invisible (e.g., S = 1 interstitials[62]), yet still contribute magnetic noise to NV-center decoherence.

Subsequent annealing at 800 °C yielded only minor additional recovery of NV coherence with $T_2 = 90$ μs. In this temperature range, vacancies are expected to become mobile and begin to aggregate into divacancies and higher vacancy complexes[44]. This trend is confirmed experimentally by the pronounced reduction of GR1 luminescence in the PL spectra (see Section B), indicating the removal or transformation of isolated $V^0$ centers. In the DEER spectra, the X resonance decreases from ~1.0 ppm to ~0.7 ppm, while $T_2$ remains nearly unchanged, suggesting that the reduction in isolated $V^-$ is largely compensated by the formation of divacancy-type defects. The divacancy ($V_2^-$) coupling constant c was extracted from the 1000 °C data, where the defect spectrum is dominated by $V_2^-$ and the model reproduces the partially recovered $T_2$ values: $c = \frac{1/T_2 - 1/T_{2,(13C)} - a[P1]}{[V_2^-]} \approx (5.0 \pm 1.7) \times 10^{-3} (\mathrm{ppm} \cdot \mu s)^{-1}$.

The most substantial recovery of long coherence times occurs only after high-temperature treatment above 1200 °C, where $T_2$ approaches ~120 μs and the residual $S = 1/2$ X-signal is reduced to ~0.1 ppm. This behavior is consistent with the progressive conversion of vacancy clusters into more benign complexes at elevated temperatures. Previous work[51] has already shown that $T_2$ does not fully return to its as-grown value after irradiation and annealing, in agreement with our observations.

The full temperature-dependent evolution of the NV coherence can then be described by a universal decoherence model (Figure 5g):

$$\frac{1}{T_2} = a[P1] + b[V^-] + c[V_2{}^-] + d[I] + \frac{1}{T_{2,(13C)}}. \quad (13)$$

The resulting relative strengths $a: b: c: d \approx 1: 2.73: 0.91: 2.18$ capture the progressive evolution of the spin bath from isolated nitrogen spins to complex vacancy and interstitial aggregates and form the quantitative basis of the universal model used in Figure 5g.

### D. Hydrogen-related side resonances and composite defect modeling

While the main DEER spectra captured the dominant X-center response, closer inspection revealed additional side-resonances in the vicinity of the main peak. These features, likely arising from hyperfine-coupled hydrogen-related defects, were reproducible across all laboratory-grown samples (S1-S3) and indicated the presence of secondary spin species not accounted for by the primary X-center model. To clarify their origin, we performed high-resolution DEER measurements with 0.2 MHz steps with finer frequency steps(Figure 6a). These features were absent (Figure 6b) in the reference diamond (S4), confirming that they are not intrinsic to the P1 or vacancy centers.

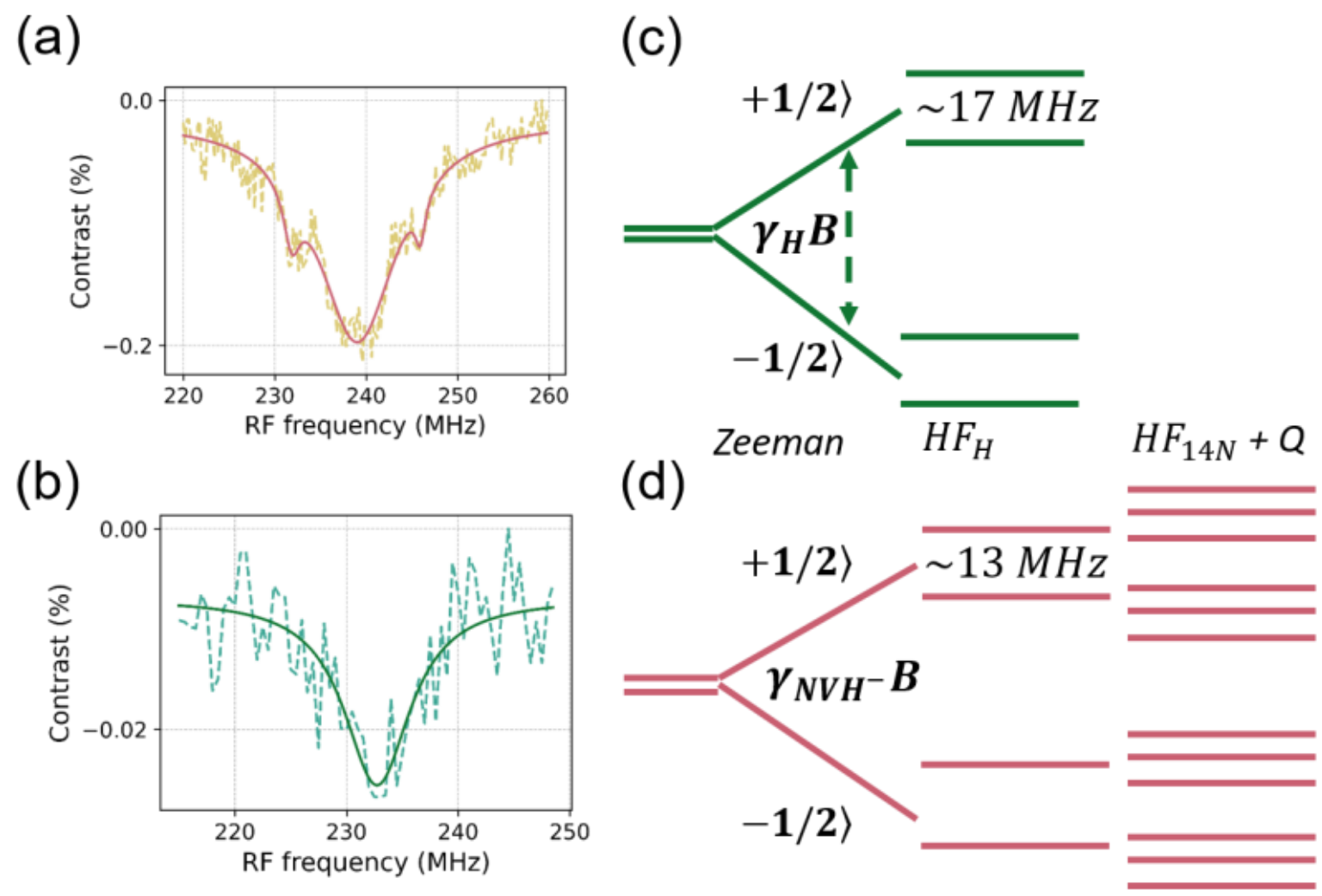

**Figure 6.** H-related defects investigation. (a) Zoom of X-resonance in the DEER spectrum of the as-irradiated S1(L4) sample. Solid line shows the Lorenz fit with 3 frequencies. (b) Zoom of X resonance in the DEER spectrum of as-irradiated reference sample S4 was fitted just by the Lorenz function. The magnetic field was different in the two cases. (c) Energy levels of the substitutional hydrogen in diamond. (d) Energy levels of the $NVH^-$center.

This difference is consistent with the fact that our CVD-grown samples already exhibited weak additional signals in their pre-irradiation spectra, likely originating from hydrogen-related defects incorporated during growth. To identify the nature of these side-resonances, we compared their hyperfine splitting with known hydrogen-related centers in diamond. Two candidates fit the observed frequency ranges: substitutional hydrogen ($H_S$) that was previously observed in EPR spectrum[33] (Figure 6c) and the $NVH^-$ defect that was previously observed in DEER spectrum[29] (Figure 6d). Both are $S = ½$ systems that can overlap spectrally with the vacancy ($V^-$) resonance and therefore produce composite features in DEER spectra.

Among all samples, S2 exhibited the strongest side-resonances, consistent with its unusually high hydrogen-to-methane ratio (99:1) during growth, which suggests enhanced hydrogen incorporation. This interpretation is supported by earlier studies[37,63], which demonstrated that increasing the impurity-to-methane ratio (in their case nitrogen) significantly enhances impurity incorporation in CVD diamond. By analogy, the pronounced hydrogen-related features observed in S2 are plausibly attributed to more efficient hydrogen uptake under its growth conditions. It should be noted that in other samples (S1, S3) we found mainly NVH defect, but could not determine their concentration from the growth at different temperatures due to the fact the detected resonances were not well pronounced to allow for quantitative analysis.

To examine these features in detail, we recorded fine-step DEER scans at five different positions on the S2 surface (Figure 7).

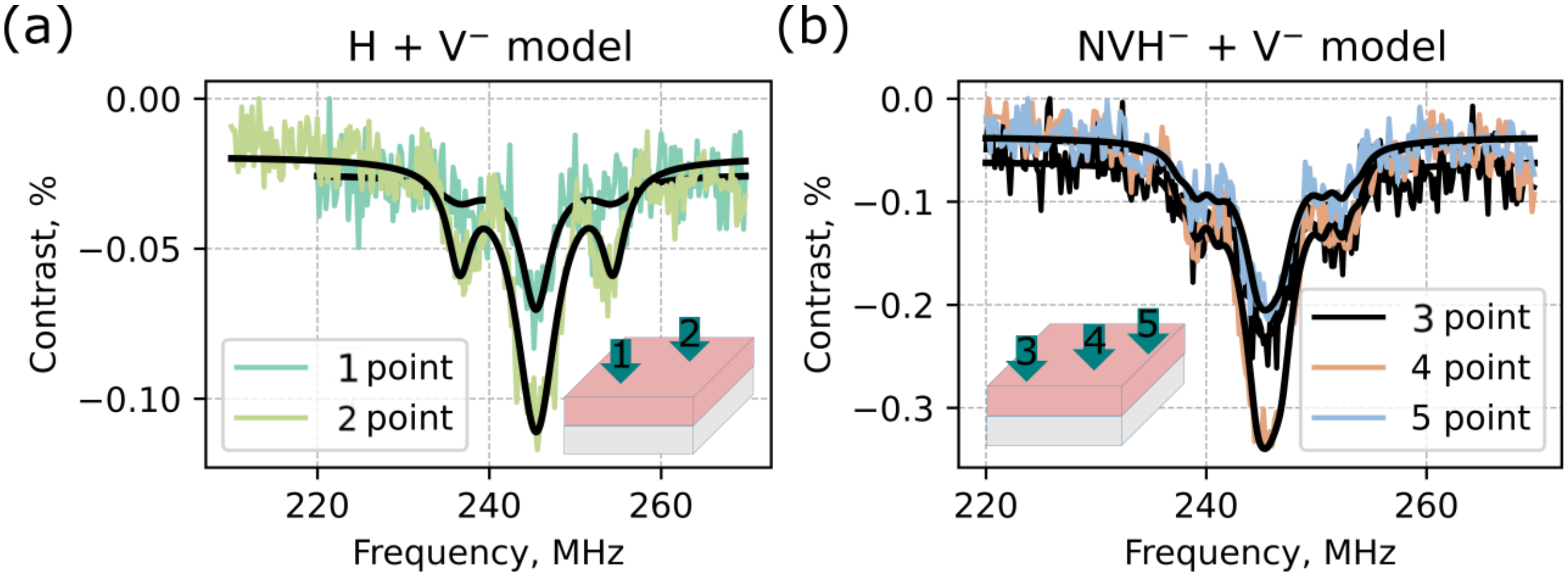

**Figure 7.** S2 sample's spectra of X-defect with side details. (a) Points that were fitted with the substitutional hydrogen + vacancy model; (b) Points that were fitted by $NVH^-$+ vacancy model; for the sample after the irradiation.

All spectra displayed side-peaks around the main X-line, but their splitting patterns varied slightly between spots. Two distinct behaviors were identified: doublets with ≈ 17 MHz hyperfine separation, and doublets with ≈ 13 MHz hyperfine separation.

Spectral modeling showed that the first group is perfectly reproduced by the composite substitutional-hydrogen + $V^-$ model $[H_H + H_{V^-}]$ (Figure 7a), while the second matches the $NVH^-$ + $V^-$ model $[H_{NVH^-} + H_{V^-}]$ (Figure 7b). The fits used literature hyperfine parameters [29,33] and treated only the magnetic-field alignment and overall amplitude as free parameters. The Hamiltonian of the substitutional hydrogen was taken from[33]:

$$H_H/h = \frac{g_e\mu_B}{h}\vec{B}g\vec{S} + \vec{S}A\vec{I}, \tag{14}$$

where $A(H)_{\parallel} = 17.9$ MHz, $A(H)_{\perp} = -2.2$ MHz, g = 2.0028. We used this Hamiltonian in model $[H_H + H_{V^-}]$ to explain spectra in some points. The Hamiltonian[29] for $NVH^-$ center:

$$H_{NVH^-}/h = \frac{g_e\mu_B}{h}\vec{B}g\vec{S} + \vec{S}A(H)\vec{I} + \vec{S}A(N)\vec{I}, \tag{15}$$

where $A(N)_{\parallel} = 2.94$ MHz, $A(N)_{\perp} = 3.1$ MHz, $A(H)_{\parallel} = 13.69$ MHz, $A(H)_{\perp} = -9.05$ MHz, g = 2.0024. In our sample dominant isotope is $^{14}N$ so defect $NVH^-$ has $S = ½$ , $I(H) = ½$ and $I(N) = 1$. $H_{NVH^-}$ was also used for explanation of sides in some of our samples. This Hamiltonian accounts for the double hyperfine splitting arising from both nuclei, however in our spectra we could not resolve $^{14}N$ hyperfine splitting.

With these two Hamiltonians, we achieved excellent agreement with the experimental data: two of the five S2 spectra were best fitted with the $[H_H + H_{V^-}]$ model, while three matched the $[H_{NVH^-} + H_{V^-}]$ model. Even though $A(H)_{\parallel}$ differs only by a few MHz between these two cases, our spectral resolution is sufficient to distinguish this (see Supplementary Materials). Hence, both hydrogen-related defects are present in our diamonds and interact with NV centers in diamond due dipole-dipole coupling. The absence of similar features in the reference sample supports the conclusion that these species originate from hydrogen incorporation during CVD growth rather than from post-irradiation processes. Together, these results show that the spin environment in hydrogen-rich CVD diamonds is more complex than in nominally pure materials, containing at least two distinct paramagnetic hydrogen species: consistent with substitutional hydrogen and $NVH^-$, that contribute to the DEER signal.

## III. METHODS

### *1. CVD Diamond Reactor and Growth Conditions*

Diamond layers were synthesized using a commercial PE-CVD system (ASTeX PDS17). Substrate temperature was monitored via a disappearing-filament pyrometer and verified by an external IR pyrometer. High-purity process gases were used: hydrogen ($H_2$, 6N–9N purity) and methane ($CH_4$, 6N–9N purity) served as the main reactants, and nitrogen was introduced through a 1% $N_2 : H_2$ premix. All flow values reported in this work have been converted to equivalent pure $N_2$ flow rates. Gas delivery was controlled by standard mass-flow controllers. The reactor operated at pressures of 100–150 Torr depending on the layer type (buffer or doped). We estimated the total depth of as-grown multilayer diamonds as also 41.2 μm with Mutitoyo Linear Gauge (LGK-0110). Individual layer's depth might be estimated by growth time (12 hours for the total diamond, and 0.5 hour for each nitrogen-doped layer, 2.5 hour for each buffer layer) that leads to 1.8 μm for each nitrogen-doped layer and 8.6 μm for each buffer layer.

Details of growth conditions for each layer are provided and discussed in Section A.

### *2. Annealing chamber*

A home-built high-vacuum annealing chamber was used for the post-irradiation thermal treatment of sample S4. The chamber employs a CeraQuest ceramic heater (PBN/PG/PBN stack) powered by an external supply. Temperature was monitored using an IRIS PI1MLO41T1800 infrared pyrometer. Each annealing step was performed for 2h at 650 °C, 850 °C, 1000 °C, and 1200 °C. After each step, the sample was re-characterized by DEER, PL, and Hahn-echo measurements to track the evolution of defect populations and NV spin properties.

### *3. Optical setup*

DEER measurements were performed using a custom-built confocal microscope (Figure 8), which allows for microwave (MW) application. The setup enables precise optical excitation and readout of $NV^-$ centers, along with MW control of their spin states. A static magnetic field of approximately 80-110 Gauss was applied using a permanent magnet, aligned along one of the four possible $NV^-$ center orientations in the diamond.

Optical excitation was provided by a 532 nm laser (HÜBNER Photonics), typically operated at ~6 mW before the objective lens (Olympus NA=0.95). The laser beam was focused onto the diamond sample, and fluorescence from the $NV^-$ centers was collected and directed to a single-photon detector (Excelitas Technologies/SPCM-AQRH-14) for signal acquisition.

Microwave pulses were generated by an arbitrary waveform generator (Keysight/M8195A) and amplified using an amplifier (Mini-circuits/ ZHL-272+). The pulses, typically ~150 ns in duration, were delivered to the diamond via a thin-wire antenna placed close to the sample surface. All experiments were conducted at room temperature under ambient conditions.

Photoluminescence (PL) measurements were performed using a spectrometer (Andor/ Kymera 193i-A) integrated into the same confocal setup described above. Optical excitation was provided by a 532 nm laser, allowing the detection of photoluminescence from optically active defects in diamond with zero-phonon line (ZPL) wavelengths longer than 532 nm.

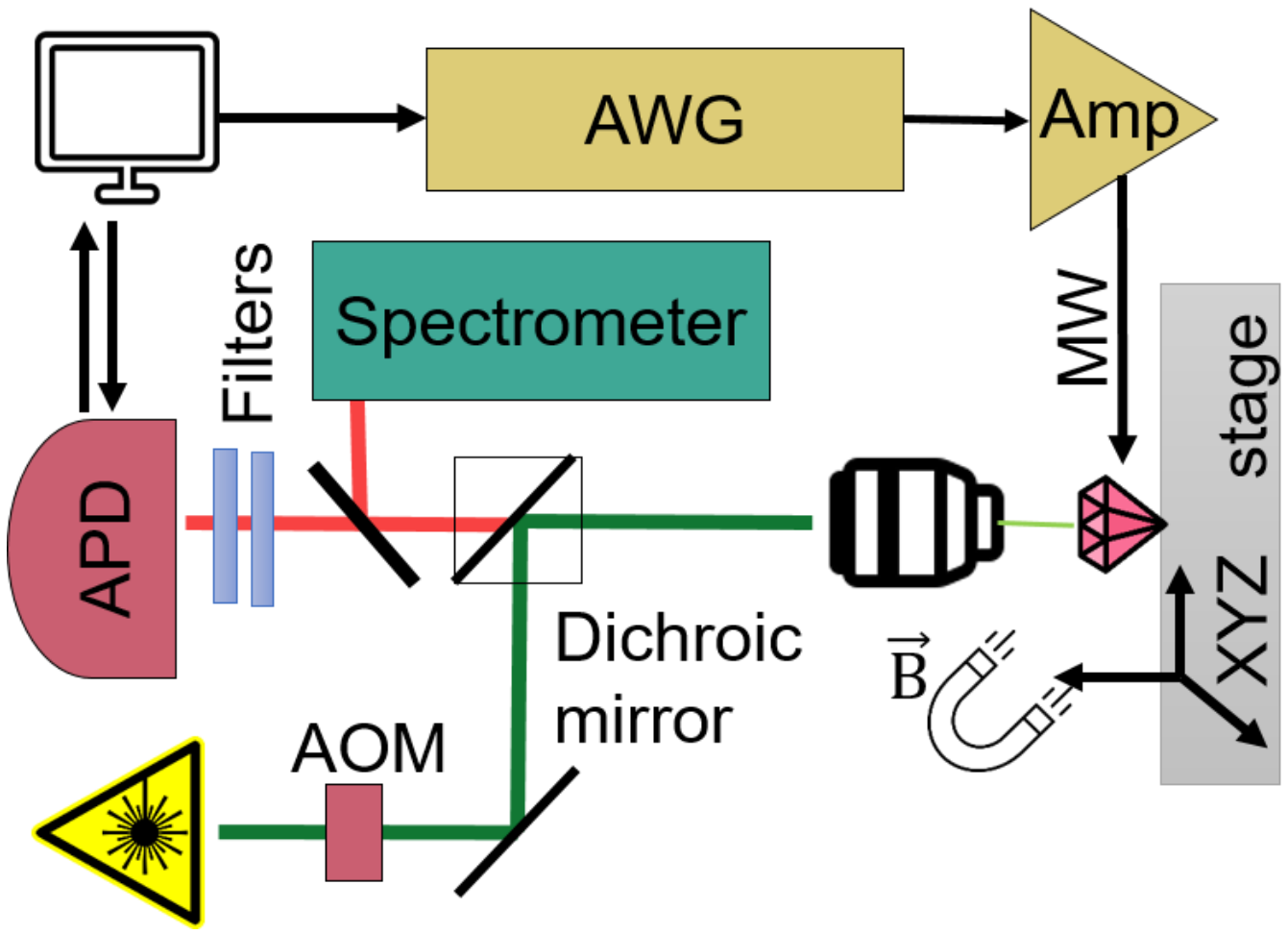

**Figure 8.** Schematic of the confocal microscope used for DEER and PL measurements, showing laser excitation and fluorescence collection paths, microwave delivery via a wire antenna, and the applied static magnetic field.

### *4. DEER spectra resolution*

In the pulsed NV-DEER measurements, the NV center (spin A) was operated in a Hahn-echo sequence, while a rectangular microwave π pulse was applied to the target spin ensemble (spin B) at the center of the echo. The DEER spectrum was obtained by scanning the microwave frequency of spin B with a step size of 0.2 MHz and recording the corresponding change in the NV echo amplitude.

The fundamental sensor-limited frequency scale is set by the NV Hahn-echo coherence time $T_2$. For $T_2 = 2.5 - 150$ μs, the corresponding Fourier-limited frequency scale $1/T_2$ is 0.4–0.007 MHz, which is substantially smaller than the MHz-wide spectral features observed in this work and therefore does not limit the experimental linewidth.

The spectral selectivity of the driven spin ensemble is determined by the finite inversion profile of the rectangular $\pi$ pulse. For $t_\pi$=200 ns, the corresponding Rabi frequency is $f = 1/(2t_\pi) =$ 2.5 MHz. This sets the characteristic frequency scale over which efficient inversion occurs. Consequently, the experimentally observed DEER linewidths arise from the convolution of the intrinsic spectral distribution of the spin ensemble (including inhomogeneous broadening and dipolar coupling distributions) with this finite inversion selectivity. In our measurements, the fitted linewidths (FWHM ≈ 3-4 MHz) are therefore consistent with intrinsic broadening of the spin ensemble combined with the finite Rabi-limited excitation bandwidth, rather than being limited by the NV coherence time or the discrete frequency sampling.

## IV. CONCLUSION

In this work, we presented an extensive investigation, aimed at providing a full analysis of the formation, transformation, and spin dynamics of defects in CVD-grown diamond at all stages of processing: from controlled growth to electron irradiation and stepwise annealing.

The goal of this work is to advance the current understanding of CVD diamond optimization for NV-based quantum sensing and to provide the wider community with an extended spin-bath model as an alternative to the widely accepted picture in which the $T_2$ spin coherence depends primarily on the substitutional nitrogen, or P1-center, spin bath used in previous works. In particular, we identify other dominant paramagnetic defects that arise during CVD diamond growth and post-processing, especially electron-beam irradiation and annealing. With a concentration uncertainty of about 10%, which we believe is low enough to perform qualitative and semi-quantitative studies, we pinpoint the defects influencing the $T_2$ times and discuss possible mechanisms behind their formation and transformation. This information is important for quantum devices such as magnetometers and for a wide range of sensing applications.

By systematically varying the growth conditions, we synthesized a set of multilayer and monolayer samples with nitrogen concentrations ranging from 1 to 40 ppm. We then used DEER and PL spectroscopy to identify and quantify their paramagnetic and optical defects.

Before irradiation, the P1 centers were the dominant spin bath, and the coherence time $T_2$ followed a well-known $T_2 \propto \frac{1}{[P1]}$ dependence. After irradiation, new paramagnetic centers appeared, collectively denoted as X-centers, whose DEER spectra revealed an $S = ½$ character and a g-factor consistent with vacancy-related defects. Annealing experiments showed that these X-centers evolve through successive stages: initially a mixture of $V^-$ and interstitial-related defects, which disappear near 650 °C, followed by divacancies that persist to ~1000 °C and finally almost disappear at 1200 °C. This sequence provides a microscopic view of how the spin bath evolves during defect engineering and affects NV coherence. A universal decoherence model was developed, including separate coupling constants for P1, $V^-$ and divacancy baths. The extracted spin coherence decay coefficients reveal that interstitial-related defects couple to NV spins even more strongly than isolated vacancies, explaining the sharp $T_2$ degradation observed immediately after irradiation.

In addition to vacancy-type centers, hydrogen-related defects such as consistent with substitutional hydrogen defect and $NVH^-$ were detected through high-resolution DEER spectroscopy, and their spectra were reproduced using multi-spin Hamiltonian modeling. Their presence correlates with hydrogen-rich growth conditions and introduces local spin noise that can further influence NV spin properties.

Overall, this work clarifies how paramagnetic defects form, transform, and interact in CVD diamond under irradiation and annealing. By combining nanoscale spin-resonance spectroscopy with optical and structural analysis, we demonstrate that NV-based DEER measurements can provide nanoscale access to hidden defects in diamond, even when they are invisible to conventional optical methods. This approach enables direct assessment of defect populations relevant for quantum sensing, magnetometry, and spin-based devices. Our results highlight that controlling the balance between P1, vacancy, and hydrogen-related defects, rather than simply maximizing NV density, is the key to achieving long spin coherence in engineered diamond systems.

While the present study focuses on bulk CVD-grown material, some of the vacancy- and hydrogen-related defects identified here are expected to also occur in nitrogen-implanted CVD diamond with shallow NV systems; however, the additional influence of implantation damage and surface-induced effects introduces a distinct spin environment, making this regime a natural subject for future investigations using the NV-DEER methodology demonstrated here.

## VI. ACKNOWLEDGMENTS

**Funding:** European Union's Horizon Europe—The EU research &innovation programme under the Grant Agreement number 101113901 and 101113983 (QU-Test and QU-Pilot projects), Grant agreement 101189875 (ACDC_Q), Grant Agreement number 101135699 (SPINUS), Grant

Agreement number 101135359 (C-QUENS), Excellent of Science CHEQS, and Strategic Basic Research, BEQNET of the Flemish Scientific Council (FWO). Fonds Wetenschappelijk Onderzoek (I011520N, G0D1721N, S008323N, 40007526) and Grantová Agentura České Republiky (24-12984S).

**Author contributions:**

Conceptualization: OR, MN
Methodology: OR, JP, DC
Investigation: OR, MP, RV
Visualization: OR
Supervision: MN
Writing—original draft: OR, JP, MN
Writing—review & editing: MP, JP, EB, MN

**Competing interests:** Authors declare that they have no competing interests.

**Materials & Correspondence**: Correspondence and requests for materials should be addressed to Milos Nesladek (email: milos.nesladek@uhasselt.be) or Olga Rubinas (email: olga.rubinas@uhasselt.be).